\documentstyle[times,pramana,epsf,psfig,floats]{ias}
\begin{document}
\mark{{Study of isomers..$^{132}Sn$...}{S. Sarkar, M. Saha Sarkar}}
\title{
Study of isomers in neutron-rich magic $^{132}$Sn  region}
\author{S. Sarkar}
\address{Department of Physics, The University of Burdwan, Golapbag, 
Burdwan -713104}
\author{M. Saha Sarkar}
\address{Saha Institute of Nuclear Physics, 1/AF, Bidhannagar, Kolkata - 700064}
\abstract{ Shell  model calculations  have  been done  for
interpreting  two  representative  isomeric states  in  neutron  rich $^{132}Sn$
region using SMN and SMPN Hamiltonians. They are, (i) 2.91 min  isomer in  $^{138}Cs$ 
and (ii) the 0.57 $\mu s$ isomer in $^{136}$Sb    nuclei.  The results  are
compared  with  those obtained  with  KH5082 and  CW5082  Hamiltonians. How  the
results clearly distinguish  the most appropriate interaction  has
been discussed. }
 
\maketitle

\section{Introduction}
Few-valence-particle  neutron- rich nuclei above the doubly magic $^{132}$Sn are
of recent interest. It  is well known that  after $^{16}$O, the strongest  shell
closure occurs at  $^{132}$Sn, and nuclei with  few  valence particles in  the
proton $\pi$(gdsh) and neutron  $\nu$(hfpi) orbitals above the  $^{132}$Sn core
are appropriate  systems inviting  applications of  the spherical  shell model.  The
region closely resembles the region above the $^{208}$Pb core. But nuclei  above
the $^{132}$Sn  core, particularly  the isotopes  of Sn, Sb, Te, I,  Xe and  Cs are
comparatively  neutron-rich and close to the dripline. For example,  the
last stable  isotope of  Sn is  $^{124}$Sn and  therefore the  Sn isotopes  with
A=134-138  are already  more than  10 neutrons away from  the line  of
stability. Study  of these  very n-rich  nuclei both  experimentally as  well as
theoretically  are important not only for the nuclear structure, such as for the
knowledge of emperical N-N  interaction in the neutron-rich  exotic environment,
but also for the applications in the astrophysical r-process model calculations.

These nuclei are located far from  the line of stability and are  very difficult
to produce.  Thermal -  neutron -  induced fission  and the  spontaneous fission
sources  of   $^{252}Cf$  and           $^{248}Cm$  initially  played  the  most
important  role  in  populating these  nuclei.  More  recently, deep  inelastic,
fragmentation and fission  at intermediate and  relativistic energies have  also
been used. Different complementary experimental techniques are used to study the
nuclear  structure  of these  neutron-rich  nuclei. Among  them  the search  for
microsecond  isomers and  the study  of their  decay schemes  are very  powerful
tools, and in many cases it is the only way to get nuclear structure information
for nuclei very far from  the stability line. Specially, the  microsecond isomer
spectroscopy is complementary to prompt gamma measurements using large  detector
arrays or 
$\beta$ decay experiments as discussed recently in Ref \cite{Pin:1}.

Near the  doubly magic nuclei, isomers, specially the microsecond ones  are very
abundant  and  these isomers  are  generally yrast  traps  carrying considerable
amounts of angular momentum. The
isomers may be studied  using recoil-fragment spectrometers which  are efficient
for the selection of the reaction  products. The detection is based on  event-by
-event time  correlation between  the fragments  and the  delayed gamma -rays or
conversion  electrons  de-exciting  the  isomers.  Recoil-fragment  spectrometer
allows  one to detect the microsecond isomers produced at a very low rate.  This
is  the  most frequently  used  technique in  recent  experiments. The  measured
energies and  half-lives of  the isomeric  transitions allow  one to  deduce the
electromagnetic transition rates which are important to test theoretical models.

In the present work shell model calculations have been done for interpreting two
representative  isomeric  states in  this  mass region.  They  are, (i) 2.91  min
isomeric  state in  $^{138}Cs$ \cite{Bnl:1}  nuclei and  (ii) the  0.57 $\mu  s$
isomer in  $^{136}$Sb \cite{Min:1}.  It is  shown clearly,  how theoretical
prediction of the  halflives of the  isomeric states sensitively  depends on
the  parametrization  of  the  shell  model  Hamiltonian  and  how  the  results
distinguish the most appropriate interaction.

\section{Shell model valence space and the (1+2)-body Hamiltonians }

The  shell  model valence  space  considered for  this  calculation consists  of
$\pi(1g_{7/2}$, $2d_{5/2}$, $2d_{3/2}$, $3s_{1/2}$  and  $1h_{11/2})$  proton orbitals  and
$\nu~(1h_{9/2}$, $2f_{7/2}$, $2f_{5/2}$,   $3p_{3/2}$, $3p_{1/2}$   and   $1i_{13/2})$  neutron
orbitals with $^{132}$Sn as the inert  core. As mentioned above, because of  its
resemblence with the  valence space above the  $^{208}$Pb core, effective  interaction
derived  for  the  well studied  Pb-region  has  been used  in  the  shell model
calculations for nuclei  in the $^{132}$Sn  region by the  appropriate scaling.
KH5082  is one  such a  Hamiltonian obtained  from the  Pb-region by  $A^{1/3}$
scaling  of  the two-body  matrix  elements (tbme)  to  take in  to  account the
different sizes  of the  nuclei in  the two  regions. With  the availability  of
experimental data on the binding  energies and low-lying spectra for  $^{134}$Sb
and $^{134}$Te, which provided information on the n-p and  p-p tbmes 
respectively, the KH5082 was  modified by  Chou
and Warburton to obtain CW5082 (1+2)-body Hamiltonian \cite{War:1}. In the recent years improved
experimental data on the binding energies and low-lying spectra for A=134, Sb
and Te have been available. It may be  noted that now
data  on  $^{134}$Sn  have also become  available  and  these  can  provide valuable
information on n-n tbmes for this  n-rich region. With these new data  on A=134,
Sn and Sb, further modification of  the CW5082 Hamiltonian has been done  recently
\cite{Sar:1} to obtain SMN  Hamiltonian. $^{134}$Te which  provides proton-proton tbmes,  has
been studied extensively. It is expected that the inclusion of new  experimental
data would further improve the SMN interaction. We have initiated this effort by
changing  only four  proton-proton tbmes  of the  SMN interaction.  The binding
energies of  $^{134}$Te with  respect to  $^{132}$Sn core  (-20.56 MeV)  and its
three low-lying excited levels  with $J^{\pi}$ = $2^+$, $4^+$, $6^+$  at energies
1279,1576 and 1692 keV, respectively, predominantly from the  $\pi(1g_{7/2})^2$
multiplet, have been used to  modify only four relevant proton-proton  tbmes and
SMN was renamed as SMPN.  These two  empirical Hamiltonians  have been  found to  be
remarkably successful in predicting  binding energies,  low-lying spectra  and
electromagnetic transition probabilities for  the nuclei in the  range A=134-138
and $50\leq Z\leq 55$.  In the  present work  we use  these two Hamiltonians along
with  KH5082 and  CW5082 in  the shell  model calculations  for $^{136}$Sb  and
$^{138}$Cs nuclei, employing the code OXBASH \cite{Bro:1}.

\section{Results and discussions}
\subsection{$^{138}$Cs }

Not much information is available on the $^{138}$Cs nucleus. This nucleus though
neutron-rich, is    not   very  exotic   (N/Z  =   1.51).  A   few  levels   are
observed \cite{Bnl:1}. But for most of them spin-parities are not assigned.  Only
three levels including the $3^-$ ground state and a 2.91 min $6^-$ isomeric state
have definite  spin-parity assignment.   It is  therefore of  interest to  apply
shell model with the newly constructed SMN and SMPN Hamiltonians for this region
to predict the  level structure and  calculate the half-life  of the isomer  for
comparison   with   the   measured  value.   Calculated  level   structure  and
electromagnetic transition probability may further act as guidance for  future
experiments on this nucleus.

\begin{figure}[htbp]
\psfig{figure=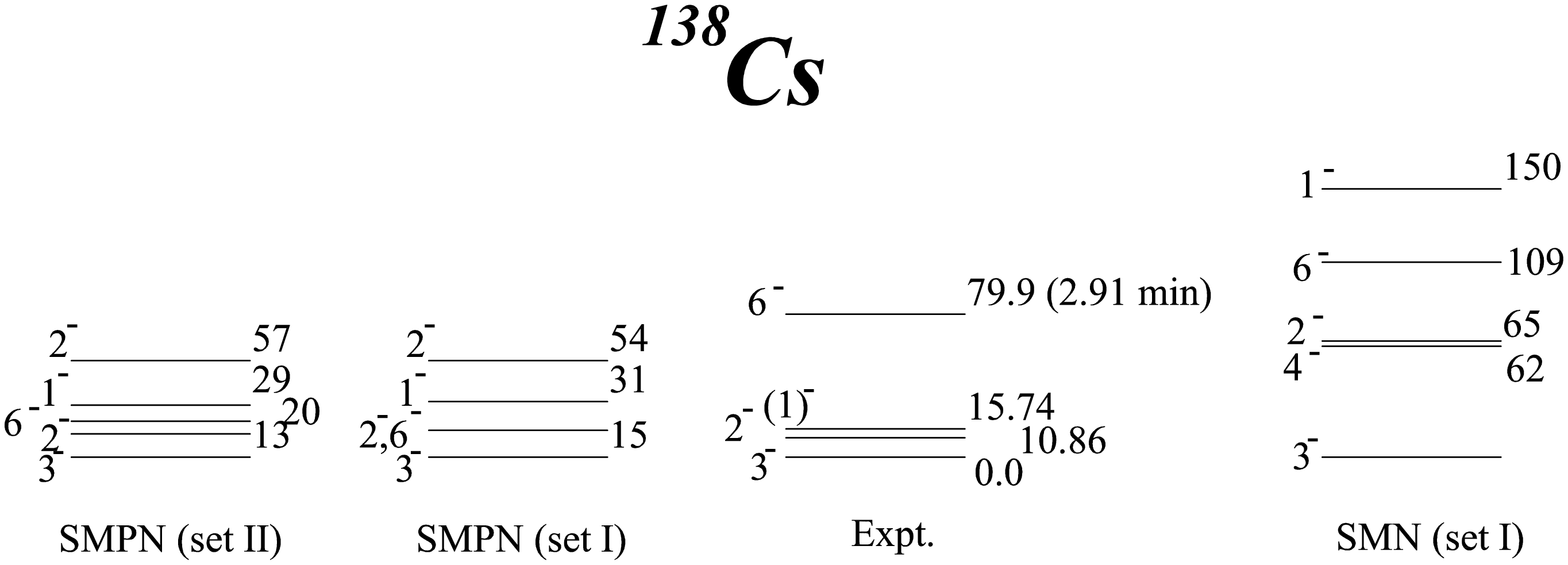,height=8.0cm,width=12.0cm,angle=90}
\caption{Calculated and experimental $^{138}Cs$ spectra.}
\end{figure}

In Fig.1, calculated  spectra with SMN  and SMPN Hamiltonians  are compared with
the  experimental  one.  In  Set  I, for  both  SMN  and  SMPN  Hamiltonians the
calculation over the valence space  mentioned above is slightly truncated.  Only
0-2 protons are  allowed to occupy  the $1h_{11/2}$ orbital, otherwise no 
truncation is  made. It correctly reproduces  a $3^-$ ground  state with both SMN 
and  SMPN Hamiltonians.

In calculation with  SMPN, the first $2^-$ excited state comes at 15 keV whereas
the  observed level  is at  10.86 keV.  The next  observed level  at 15.74  keV
excitation has been assigned $(1)^-$ and we get a $1^-$ level at 31 keV. In  the
untruncated calculation,  SMPN (Set  II), 2$^-$  comes at  13 keV  and the 6$^-$
comes  at 20  keV. the  first excited  1$^-$ now  comes at  29 keV.   The most
interesting  is  that both  the  calculations correctly  reproduce  the isomeric
nature of the $6^-$ level though  its excitation energy is underpredicted to  be
at 15 keV (SMPN Set I), and 20 keV (SMPN Set II). The measured excitation energy
of this state is  79.9 keV. The only  favourable transition is $6^-  \rightarrow
3^-$ which is either a M3 or E4 or both and consequently result in a large  half
-life for the isomeric decay.  The $4^-$,$5^-$,$7^-$ states are above  the first
$6^-$ state. For most of the yrast states from $7^-$ to $0^-$, except $1^-$  and
$6^-$, the  dominant components  of the  wave functions   are $\pi  {1g_{7/2}}^3
{2d_{5/2}}^2 \nu{2f_{7/2}}$ and  $\pi {1g_{7/2}}^5 \nu 2f_{7/2}$, the former one
is 28-38\% and the  later is 13-30\%.  We  name these two configurations  C1 and
C2. For $1^-$ the  C1  configuration is 28\%  with an admixture of  16\% of $\pi
{1g_{7/2}}^4 2d_{5/2} \nu 2f_{7/2}$.  The  wave function structure of the  $6^-$
isomeric state is, as one may expect, quite different than other yrast states.
Its dominant components  are not C1  and C2 but  $\pi {1g_{7/2}}^4 2d_{5/2}  \nu
2f_{7/2}$(55.2\%) and $\pi  {1g_{7/2}}^2 {2d_{5/2}}^3 \nu  2f_{7/2}$(16.1\%). It
should be mentioned here that the results of the untrunated calculation (Set II)
validate the results of the slightly  truncated one (Set I) by showing  that the
occupation of the intruder orbital $\pi h_{11/2}$ is small.

For the calculations with SMN interaction, all the yrast states including
$6^-$, C2 is most dominant (30-50\%) and C1 is 23-30\%. Moreover, in
calculation with  SMN, while $3^-$ is still the ground state but the 
$1^-$ state is pushed up in energy whereas $2^-$ and $4^-$ come below 
the $6^-$ level. So in this calculation $6^-$ apparently does not look
like an isomer. 

 Half-life of the  isomeric state is  then calculated with  the wavefunctions of
 both SMN and SMPN (both  Sets)  calculations. Calculated half-life for  the $6^
 - \rightarrow 4^-$ E2 transition with the wavefunctions of the SMN  Hamiltonian
 is of $\mu$s order.  The measured half-life is  2.91 min. Thus SMN  calculation
 fails to predict the longlived  isomer. For calculation with the  wavefunctions
 of the SMPN Hamiltonian, $6^- \rightarrow 3^-(gs)$  M3 transition gives a  half
 -life of about  12 min. Bare  values of the  g-factors have been  used. E4 half
 -life  for  the  same  transition is  extremely  large.  Thus  SMPN Hamiltonian
 correctly predicts a long lived isomer in $^{138}$Cs. The result also indicates
 a need  for larger  B(M3) value  implying thereby  different g-factors than the
 bare ones.

\subsection {$^{136}Sb$}
At present very little is known  about the nuclear structure of $^{136}Sb$.  The
ground state of this nucleus is argued to be 1$^-$ by Hoff et al \cite{Hof:1}. Their argument
was based  mainly on  the strong  beta transitions  of the  $^{136}Sb$
ground state to the  $I^\pi$ =  0$^+$ ground state  and the  excited 2$^+$
states  in  $^{136}Te$.  They  also used  the   analogy  of  $^{136}Sb$ with  the
$^{212}Bi$ of the $^{208}Pb$ region, the two regions being  similar.  The ground
state of $^{212}Bi$ is  I$^\pi$ = 1$^-$. The  ground state in $^{136}Sb$  arises
predominantly from  the $\pi  1g_{7/2}$ $\nu(2f_{7/2})^3$  configuration. So  it
must have negative parity. A 0$^-$ ground state would imply a highly  improbable
scenario  where  several states  below  3 MeV  in  $^{136}Te$ are  populated  by
relatively strong  first forbidden  unique beta  transitions. Hoff  et al.  also
ruled  out the possibility  of a 2$^-$ ground  state only on the  ground that it
demanded an exceptionally  fast log ft$^{1u}$  =8.2 first forbidden  unique beta
transition to  the 0$^+$  ground state  of $^{136}Te$.  They used,  in favour of
their argument, the smallness of the unique forbidden matrix elements from the beta  decay
of $^{134}Sb$ and $^{135}Sb$.  

 In our calculations  with  both  CW5082  and SMN Hamiltonians,  we get a  $2^-$
 ground  state  for $^{136}Sb$.  The   first excited  1$^-$  state comes  at  an
 excitation  energy  of  125  keV   and  47  keV,  respectively.   With   KH5082
 Hamiltonian, the ground state  comes out to be  1$^-$. The first excited  $2^-$
 state comes at 126 keV. Since KH5082 predicts incorrectly  a 1$^-$ ground state
 instead of 0$^-$ in  $^{134}Sb$, and fails badly  compared to CW5082 and  SMN,
 for  $^{135}Sb$ and other  N=84 isotones, the  prediction of a 1$^-$  ground
 state for $^{136}Sb$  by  KH5082 (and  its different variations \cite{Khn:1}),  which agrees
 with the assignment of Hoff et al., seems to be questionable.

The only information available on $^{136}$Sb are, a beta delayed neutron-emission half-life
of 0.923(14)s, a beta delayed neutron-emission probability of 16.2(32)\% per decay and a
 $\mu$s isomeric state with a half-life of 0.565(50)$\mu$s observed recently through a 
173 keV gamma transition  by Mineva et al \cite{Min:1}. They  examined several probable scenarios for this isomeric transition. The 
most probable scenario they advanced was that the isomer originated from the small energy spacing between the $6^-$ and $4^-$  levels of the $\pi(1g_{7/2})$ $\nu(2f_{7/2})^3$ multiplet. This was based on their shell model calculations with Kuo-Herling interaction (KH5082 and its variations,
like Khn \cite{Khn:1})  and the non-observation of a high-spin ($6^-$ ) beta-decaying isomer.
 According to them the observed 173 keV $\gamma$-ray was the result of E2 transition from the 
$4^-$ to $2^-$ level and the low energy $6^-$  to $4^-$ and $2^-$ to $1^-$ (ground state)
 transitions were not observed due to conversion and/or absorption in the aluminum catcher. 
In deducing their theoretical B(E2) value, for comparison with that from the measured half-life,  
they, however, did not mention the effective charges used.

Our calculation of half-life for the excited states also favour the $6^-$ level as the probable isomeric state.  It seems that a $2^-$ assignment to the ground state spin-parity makes the discussion of the isomeric transition easier, compared to the $1^-$ assumption. 
From the measured half-life of $565 \pm 50$ ns of the isomeric state one can deduce a 
value for the  
B(E2)=${{4.11}^{+0.40}}_{-0.33}$ W.u. with transition energy 47 keV (not observed) between 
$6^-$  to $4^-$ levels. This energy for the gamma-ray is obtained in our shell model 
calculation with SMN Hamiltonian. The
$4^-$ to $2^-$(ground state) transition energy 148 keV observed as 173 keV transition 
experimentally. The calculated half-life of the $4^-$ state is  much shorter compared to 565 ns. 
The effective charges for reproducing experimental half-life of the $6^-$ isomeric
state are proton effective charge = 1.00 and neutron effective charge = 
${{1.00}^{+0.06}}_{-0.05}$ 
(in unit of e), seem reasonable when compared with those for the N=84 isotones \cite{Sar:1}.

\section{Conclusion}
Only SMPN Hamiltonian in the mass region can reproduce correctly the $6^-$
isomeric state in $^{138}$Cs.  Recalling the fact  that the SMPN  differs 
from the  SMN only by  four important $\pi-\pi$ tbmes, it  shows how 
sensitively  the wavefunctions depend  on certain important tbmes. 

Thus changes in the $\pi-\pi$  tbmes in SMPN is significant and more 
$\pi-\pi$ tbmes should be modified in the light of new experimental data
to improve the  prediction for  the energy  eigenvalues of low-lying
yrast levels. Origin of the isomeric transition in $^{136}$Sb has been 
discussed and effective charges for N=85 Sb has been derived. The question
of the ground state spin of $^{136}$Sb can be resolved only through further
experimental investigation on this nucleus. Work is in progress to compare
the shell model prediction of beta-delayed n-emission half-life with the 
measured value. This will further help in inferring about the quality
of the shell model wave functions.


\begin{thebibliography}{99}
\bibitem{Pin:1}  J.A. Pinston and J. Genevey, J. Phys. G: Nucl. Part. Phys. 30 (2004) R57.
\bibitem{Bnl:1}  Data  extracted  using  the  NNDC  On-line  Data
Service from ENSDF and XUNDL databases.
\bibitem{Min:1}  M. N. Mineva et al., Euro. J. A 11 Phys. (2001) 9.
\bibitem{War:1} W.T.  Chou  and E.K. Warburton, Phys. Rev. C 45, 1720 (1992).
 \bibitem{Sar:1}  Sukhendusekhar  Sarkar,  M.  Saha Sarkar, Eur. Phys.
Jour. A {\bf 21} (2004) 61 and references therein.
\bibitem{Bro:1} B.A.Brown,  A.Etchegoyen,  W.D.M.Rae   and N.S.Godwin, MSU-NSCL Report No.  524, 1985 (unpublished).
\bibitem{Hof:1} P. Hoff et al., Phys. Rev. C 56 (1997) 2865.
\bibitem{Khn:1} For example, new experimental single particle energies for some single 
particle orbitals with the KH5082 tbmes.
\end{thebibliography}
\end{document}